\documentclass[proof]{pasj00}
\usepackage{graphicx} 

\begin{document}
\SetRunningHead{Author(s) in page-head}{Running Head}
%\Received{}%{yyyy/mm/dd}
%\Accepted{}%{yyyy/mm/dd}
%\Published{}%{yyyy/mm/dd}

\title{Suzaku View of the Neutron Star in the\\
Dipping Source 4U 1822$-$37}

%%% begin:list of authors
% Do NOT capitalize all letters in "textsc".
\author{Makoto \textsc{SASANO}\altaffilmark{1}, Kazuo \textsc{Makishima}\altaffilmark{1,2,3}, Soki \textsc{SAKURAI}\altaffilmark{1}, Zhongli \textsc{Zhang}\altaffilmark{1}, Teruaki \textsc{ENOTO}\altaffilmark{2,4}} 
%  \thanks{Example: Present Address is xxxxxxxxxx}
%\affil{Department of Physics, School of Science, The University of Tokyo, 7-3-1, Hongo, Bunkyo-ku, Tokyo 113-0033, Japan}
\altaffiltext{1}{Department of Physics, School of Science, The University of Tokyo, 7-3-1, Hongo, Bunkyo-ku, Tokyo 113-0033, Japan}
\altaffiltext{2}{Cosmic Radiation Laboratory, Institute of Physical and Chemical Research (RIKEN), Wako, Saitama 351-0198, Japan}
\altaffiltext{3}{Research Center for the Early Universe, The University of Tokyo, 7-3-1, Hongo, Bunkyo-ku, Tokyo 113-0033, Japan}
\altaffiltext{4}{NASA Goddard Space Flight Center, Astrophysics Science Division, Code 662, Greenbelt, MD 20771, USA}
\email{sasano@juno.phys.s.u-tokyo.ac.jp}

%\author{Kazuo \textsc{Makishima}}
%\affil{Department of Physics, School of Science, The University of Tokyo, 7-3-1, Hongo, Bunkyo-ku, Tokyo 113-0033, Japan}\email{bbbbb@xxx.xxx.xx.xx}
%\and
%\author{C-Firstname {\sc C-Familyname}}
%\affil{C-Address of Institute}\email{ccccc@xxx.xxx.xx.xx}
%%% end:list of authors

%%% Please use the following style in case that sorting by 
%%% affiliation is impossible. 
%
% \author{%
%   D-Firstname \textsc{D-Familyname}\altaffilmark{1}
%   E-Firstname \textsc{E-Familyname}\altaffilmark{1,2}
%   and
%   F-Firstname \textsc{F-Familyname}\altaffilmark{2}}
% \altaffiltext{1}{Address of Institute}
% \email{ddddd@xxx.xxx.xx.xx}
% \email{eeeee@xxx.xxx.xx.xx}
% \altaffiltext{2}{Address of Institute}

%% `\KeyWords{}' always has to be placed before `\maketitle'.
\KeyWords{accretion $-$ Stars:  magnetic field $-$ X-rays : binaries} %Do NOT move this preamble from here!

\maketitle

\begin{abstract}
The dipping X-ray  source 4U 1822$-$37  was observed by {\it Suzaku} on 2006 Octrober 20 for a net exposure of 37 ks.
The source was detected with the XIS at a 1-10 keV flux of 5.5$\times10^{-10}$ erg cm$^{-2}$ s$^{-1}$, and with the HXD (HXD-PIN) at a 10-50 keV flux of 8.9$\times10^{-10}$ erg cm$^{-2}$ s$^{-1}$.
%The count rates are $\sim$10 cnt s$^{-1}$ and 1.6 cnt s$^{-1}$ with the XIS and the HXD-PIN, respectively.
With HXD-PIN, the pulsation was detected at a barycentric period of 0.592437 s, and its change rate was reconfirmed as $-2.43\times$10$^{-12}$ s s$^{-1}$.
The 1-50 keV spectra of 4U 1822-37 were found to be very similar to those of Her X-1 in the slopes, cutoff and iron lines.
Three iron lines (Fe K$\alpha$, Fe XXV, and Fe XXVI) were detected, on top of a 1-50 keV continuum that is described by an NPEX model plus a soft blackbody.
In addition, a cyclotron resonance scattering feature was detected significantly ($>$99\% confidence), at an energy of 33$\pm$2 keV with a depth of 0.4$^{+0.6}_{-0.3}$.
Therefore, the neutron star in this source is concluded to have a strong magnetic field of 2.8$\times10^{12}$ G.
Further assuming that the source has a relatively high intrinsic luminosity of several times 10$^{37}$ erg s$^{-1}$, 
its spectral and timing properties are consistently explained.
%By the spectral fitting, the continuum of spectra are reproduced by black body and NPEX with cyclotron resonance scattering feature and three iron lines, Fe K$\alpha$, Fe XXV and Fe XXVI, are detected.
%The significance of the CRSF is sured over 99\% by both F-test and CRSF depth.
%From the relation between the cutoff energy and the photon index, 4U 1822-37 is classified as a strong magnetic field neutron star.
%By the energy of cyclotron resonance scattering feature, the magnetic field strength of 4U 1822-37 was measured, $3\times10^{12}$ G.
%With the strong magnetic field, the spectral shape of the continuum, the relation between pulse period and its change rate, and the iron line features are clearly interpreted.
%It is considered that typical LMXBs have weak magnetic field strength ($< 10^{10}$ G).
%LMXBs which have strong magnetic field strength ($\sim 10^{12}$ G) are rarely discovered.

\end{abstract}

\section{Introduction}

Most of low-mass X-ray binaries (LMXBs), namely X-ray emitting binaries with Roche-lobe filling low mass stars, are considered to involve neutron stars (NSs) with low magnetic field strengths, $B$ $<$10$^{10}$ G. 
They make a contrast to high-mass X-ray binaries (HMXBs), mostly containing NSs with high magnetic field strengths ($B$ $\sim$10$^{12}$ G) which capture stellar winds from their companions.
In fact, among $\sim$50 known LMXBs, only three are known to have NSs with strong magnetic fields; Her X-1, 4U 1626$-$67 and GX 1+4 \citep{1989PASJ...41....1N}.
These distinct combinations of the mass-donating and mass accreting components may be generally interpreted as population effects,
that older NSs have weaker fields.
However, it is not necessarily clear whether NSs gradually lose their magnetic fields (e.g., \cite{1992ARA&A..30..143C}).
Then, it is worth while searching for other LMXBs that involve NSs with strong magnetic fields.

Evidently, the magnetic field affects accretion mechanisms in binaries. 
When the NS has strong magnetic fields, the accreting matter is funneled onto the two magnetic poles,
leading to strong pulsations, and the production of very hard X-ray spectra which is often accompanied by cyclotron resonance scattering features (CRSFs; \cite{1999ApJ...525..978M}).
If, in contrast, the NS is weakly magnetized, an accretion disk is considered to extend down to vicinity of the NS, 
and the emergent X-ray spectra will exhibit characteristic bimodel behavior between so-called soft state and hard state (e.g. \cite{2007ApJ...667.1073L,2012PASJ...64...72S}).
These spectral properties, together with the presence/absence of X-ray pulsations, will conversely allow us to tell whether the NS in a mass exchanging binary is strongly or weakly magnetized.
%accretion gases which were not squeezed by magnetic field fall to all or wide area of the neutron star.
%We do not understand effects of magnetic field enough because there are only a few LMXBs which have strong magnetic fields. 
%No strongly magnetized NSs lurk the known LMXBs any more?

%Typical LMXBs show state transitions like black-hole binaries, clearly different spectra against HMXB's ones and no clear pulsation by the weak magnetic field.
%By the weak magnetic field strength, it is considered that an accretion disk go near a neutron star and 
%accretion gases which were not squeezed by magnetic field fall to all or wide area of neutron star.
%It is also considered that X-ray bursts of LMXBs are occurred by the weak magnetic field.
%In a case of the strong one, hydrogen or helium gas may burn in an denser accretion column which is squeezed by the magnetic field.
%We try to search LMXBs which have strong magnetic field.

%about 4U1822-37
In an attempt of looking for LMXBs that involve strongly magnetized NSs, the present paper focuses on the dipping X-ray binary 4U 1822-37, 
located at an estimated distance of 2.5 kpc \citep{1982MNRAS.200..793M,2003AJ....125.2163C}.
%The dips of this source were firstly reported by Mason and Cordova (1982).
The companion star in 4U 1822$-$37 is estimated to have a mass of 0.44 - 0.56 {\it M$_{\odot}$} \citep{2005ApJ...635..502M}, with $M_{\odot}$ being the Solar mass, and 
hence this system is clarified as an LMXB.
%the distance to the binary system is 2.5 kpc (Cowley et al. 2003, Mason and Cordova 1982).
%X-ray and optical light curves of this source both show dips synchronized with its orbital period, $P_{\rm orb}$ $\sim$ 5.7 hr \citep{1982MNRAS.200..793M,2010A&A...515A..44B}.
X-ray and optical light curves of this source both show intensity modulations and dips synchronize with its orbital period, 
$P_{\rm orb}$ $\sim$ 5.7 hr \citep{1982MNRAS.200..793M,2010A&A...515A..44B}. 
These effects are attributed to either gradual occultation (by the companion star) of a large X-ray scattering corona above the accretion disk (e.g., \cite{2001ApJ...557...24I}), 
or variable attenuation of direct X-rays from the NS by ionized humps on the disk (e.g., \cite{2006A&A...445..179D}).
From dip properties, the orbital inclination is constrained between 76$^{\circ}$ and 84$^{\circ}$ \citep{1989MNRAS.239..715H,2001MNRAS.320..249H}.

%X-ray pulsations with period $P_{\rm s}$ = 0.5925 s and 
%the change rate of $P_{\rm s}$, $\dot{P}_{\rm s}$=$-2.85\times 10^{-12}$ s s$^{-1}$ were also detected (Jonker et al. 2001).
%A change rate of the orbital period is also derived $\sim10^{-10}$ s$\cdot$s$^{-1}$ with many satellites observation.
%Jonker et al. 2001 reported a detection of pulsation whose period is 0.592 s and estimated magnetic field strength (10$^{8-16}$ G) with luminosity and pulse period.

\begin{figure}
  \begin{center}
    \includegraphics[width=80mm,height=70mm]{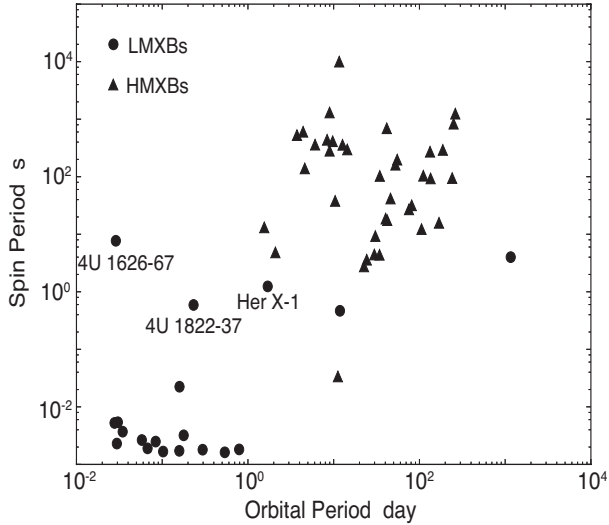}
    %%% \FigureFile(width,height){filename}
  \end{center}
  \caption{The Corbet diagram for neutron-star binaries of known rotation periods. Circles are LMXBs and triangle HMXBs.}\label{corbet_diagram}
\end{figure}

%what is the difference of 4U1822-37
Although 4U 1822$-$37 is classified as an LMXB, its X-ray properties appear rather different from those of other dipping LMXBs.
First, it shows clear X-ray pulsations with a period of $P_{\rm s}$ =  0.5924 s 
which is much longer than those of other typical LMXBs, a few milliseconds \citep{2004NuPhS.132..496W} if ever detected.
Figure \ref{corbet_diagram} is so-called Corbet diagram (\cite{1984A&A...141...91C}), which displays neutron-star binaries on a plane of orbital period and rotation period.
Like 4U 1626$-$67 and Her X-1, 4U 1822$-37$ is thus located on this plot between typical LMXBs and HMXBs.
Combining the estimated X-ray luminosity (10$^{36-38}$ erg s$^{-1}$) with $P_{\rm s}$ and its change rate, $\dot{P}_{\rm s}$ = $-2.85\times 10^{-12}$ s s$^{-1}$, 
the magnetic field strength of the NS in this source was actually estimated as $B=10^{8-16}$ G (\cite{2001ApJ...553L..43J}).
Although the error is very large, it is noteworthy that rather high field strength are allowed.

%with large error,B $\sim$ 10$^{8-16}$ G.
%The change rate of $P_{\rm s}$, $\dot{P}_{\rm s}$ was also measured $-2.85\times 10^{-12}$ s s$^{-1}$.
Second, this object exhibits much harder spectra together with a lower cutoff energy than typical LMXBs in the hard state.
From the measured cutoff energy ($\sim$ 6 keV), \citet{2000A&A...356..175P} obtained an estimate as $B$ $\sim$ 10$^{12}$ G.
Third, \citet{2010ApJ...715..947T}, using the {\it Chandra HETGS}, detected from this source narrow iron K$_{\alpha}$ and K$_{\beta}$ lines,
which are generally absent or much weaker in other LMXBs \citep{2008ApJ...674..415C}.
In the present study, we examine whether the NS in 4U 1822$-$37 has {\it B} $\sim$ 10$^{12}$ G or not.
For this purpose, it is necessary to obtain spectra with a good energy resolution and a wide energy coverage.
The fifth Japanese X-ray satellite Suzaku \citep{2007PASJ...59S...1M}, carrying onboard 
the XIS \citep{2007PASJ...59S..23K} and the HXD \citep{2007PASJ...59S..35T}, is best suited for the requirements.
We hence utilized an archival Suzaku data set of 4U 1822$-$37, and study its spectral and pulsation properties.
The mechanism of the periodic X-ray and optical dips, though interesting, is beyond the scope of the present paper.
%To study with spectral features, it is necessary to obtain spectra with good energy resolution and wide energy band.
%To achieve obtaining the spectra, The Japanese fifth satellite {\it Suzaku} (Mitsuda et al. 2007) is best one.
%Using them, we are able to obtain very wide band spectra (0.5 to 600 keV) and good energy resolution at the low energy band at the same time.
%We analysed {\it Suzaku} public data to study magnetic field strength of 4U 1822-37.

%\noindent IMPORTANT NOTICE\\
%1. ``\verb|\draft|'' creates single column and double spaces format.\\
%2. If you comment out ``\verb|\draft|'', the output will be double column
%   and single space.\\
%3. For cross-references, the use of ``\verb|\label|, \verb|\ref|, \verb|\cite|'' 
%   and the thebibliography environment is strongly recommended. \\
%4. Do NOT use ``\verb|\def|, \verb|\renewcommand|''.\\
%5. Do NOT redefine commands provided by PASJ00.cls.\\

\section{Observations and Data Reduction}
With Suzaku, 4U 1822$-$37 was observed for an exposure 37 ks on 2006 October 20 (ID 401051010).
%The observation data were a public data and ID is 401051010.
In this observation, the target was placed at the XIS nominal position.
Of the four XIS cameras, XIS0 was operated in the 1/4 window mode, and the others in the full window mode. 
The HXD was operated in the normal mode.
%The XIS has three From sided Illuminated (FI) CCD cameras (XIS0,2 and 3)  and a Back sided Illuminated (BI) one (XIS1).

Using HEAsoft ver 6.12, we analyzed the data from XIS0, XIS2, XIS3 and HXD-PIN.
In this study, we did not use XIS1 which is the back-illuminated camera, 
because it has a higher background and is subject to larger calibration uncertainties than the other XIS cameras.
As the average counts rate of the XIS, $\sim$ 10 cts s$^{-1}$, was high enough to cause pile up under the full window mode,
the XIS events were accumulated over an annular region with the inner and outer radii of 0$^{\prime}$.5 and 2$^{\prime}$.0, respectively.
%In our analysis, we chose a circular ring (inner radius is 0.5$^{\prime}$ and outer radius 2.0$^{\prime}$) to exclude brightest point. 
This allowed us to reduce pile up effects to within 1\% \citep{2012PASJ...64...53Y}.
The response and arf files of the XIS were generated using  {\tt xisrmfgen} and {\tt xissimarfgen}, respectively.
The HXD-PIN data were analyzed using a ``tuned'' non X-ray background file and the {\tt epoch2} response file, both officially released by the HXD team.
We did not use the HXD-GSO data because the source was undetectable therein.

The present Suzaku data were already analyzed by \citet{2011A&A...534A..85I}.
Combined with previous data, they determined the orbital period to be $P_{\rm orb}$ = 20054.2049 s = 5.57 hr,
but did not report on pulse detection, or spectral analysis.

\section{Results}
\subsection{Timing analysis}
\begin{figure}
  \begin{center}
    \includegraphics[width=80mm,height=80mm]{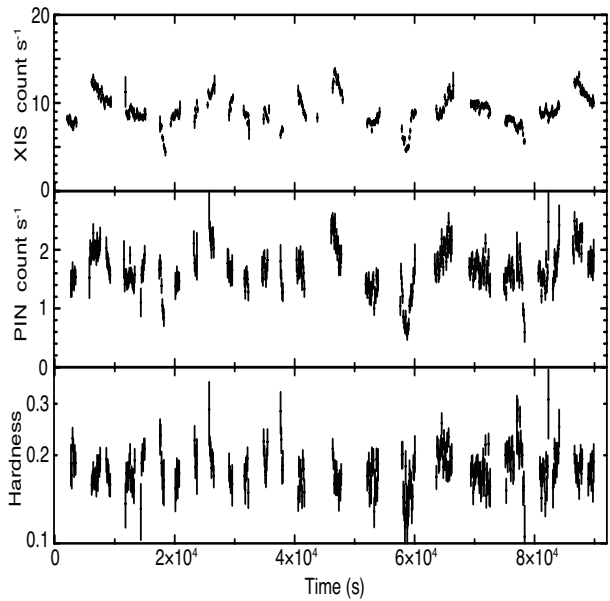}
    %%% \FigureFile(width,height){filename}
  \end{center}
  \caption{Background subtracted light curves of 4U 1822$-$37 with 128 s bins, obtained with XIS0+XIS3 (top panel; 1-10 keV) and HXD-PIN (middle panel; 15-60 keV).
The bottom panel shows the HXD-PIN vs XIS hardness ratio.}\label{lightcurve}
\end{figure}

\begin{figure}
  \begin{center}
    \includegraphics[width=80mm,height=80mm]{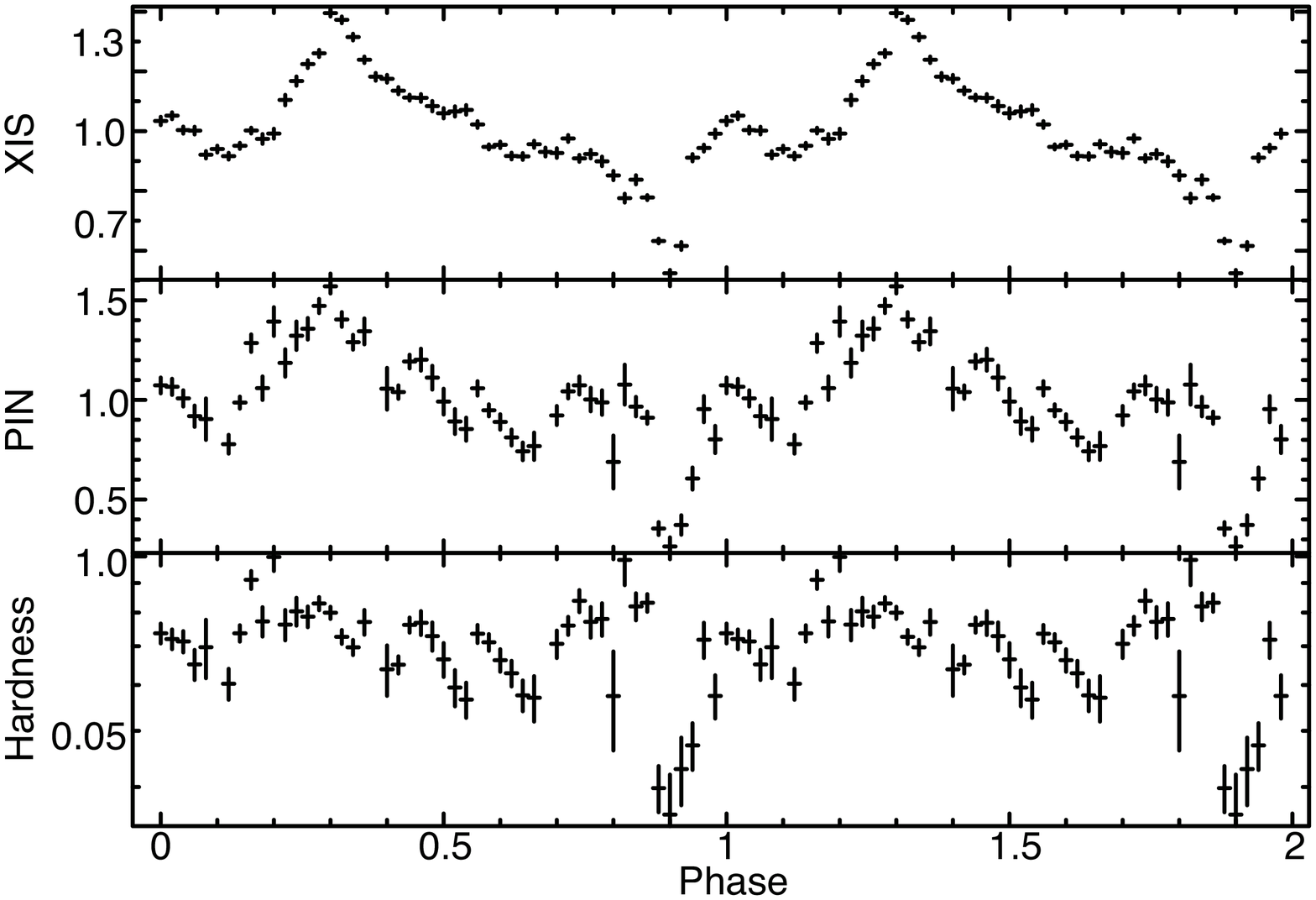}
    %%% \FigureFile(width,height){filename}
  \end{center}
  \caption{The same light curves of 4U 1822$-$37 as in figure\ref{lightcurve}, folded at the orbital period of 20054 s. The three panels correspond to those of figure \ref{lightcurve} .}\label{foldedlightcurve}
\end{figure}

Figure \ref{lightcurve} shows background-subtracted light curves of 4U 1822$-$37 obtained with 
the XIS (1 to 10 keV) and HXD-PIN (15 to 60 keV), together with the HXD-PIN vs. XIS hardness ratio.
The gross exposure of this observation ($\sim$90 ks) covered about four orbital cycles,
in which we detected about 4 dips in both the XIS and HXD-PIN bands.
However, unlike the cases of many other dipping sources (e.g. XB 1916$-$053, XB 1323$-$619 and EXO 0748$-$176),
no X-ray bursts were detected.
%However, we did not detect X-ray burst like other dipping sources (e.g. XB 1916-053, XB 1323-619 and EXO 0748-176)
The same light curves, folded at $P_{\rm orb}$=20054.2049 s, are shown in figure \ref{foldedlightcurve}.
%We folded light curves by the orbital period of 20054.2049 s (Iaria et al. 2011), and show the results in figure.\ref{foldedlightcurve}.
They are consistent with those of previous studies (e.g., \cite{2008ApJ...673.1033I,2011A&A...534A..85I}).
From the bottom panel of figure \ref{foldedlightcurve}, the hardness ratio is observed to decrease during the dips.

\begin{figure*}
  \begin{center}
    \includegraphics[width=160mm,height=80mm]{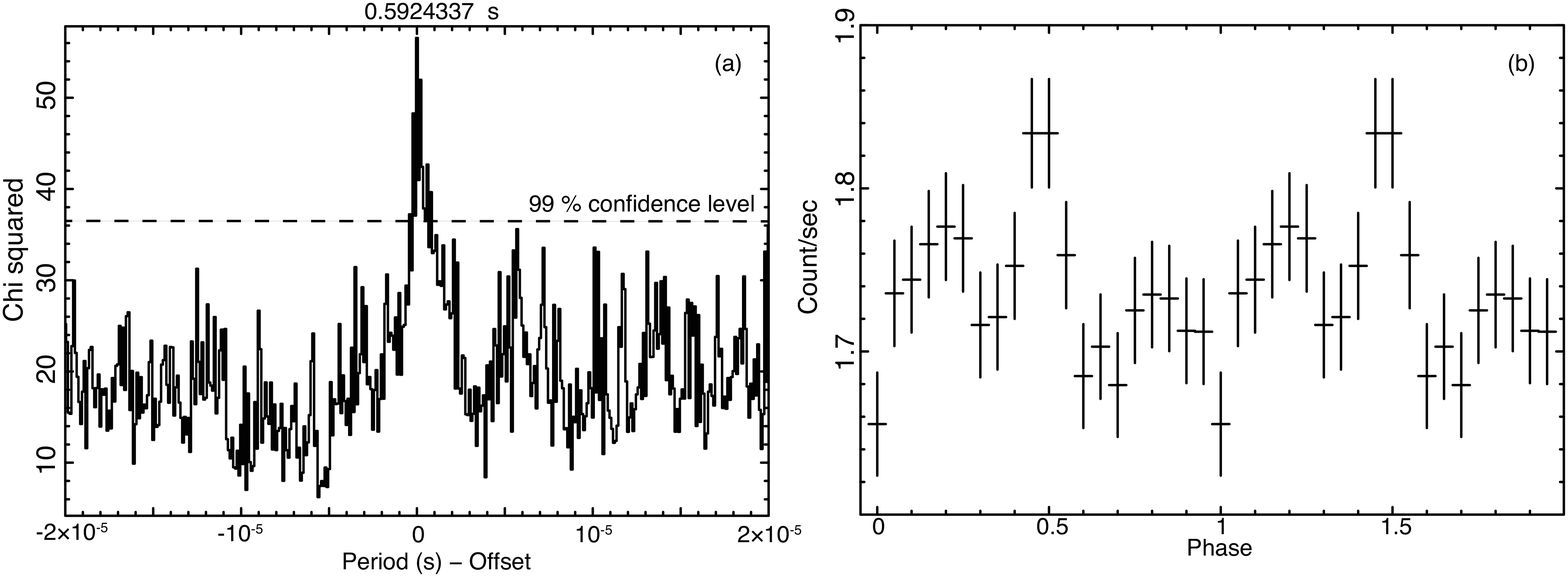}
    %%% \FigureFile(width,height){filename}
  \end{center}
  \caption{(a) : The periodogram of 4U 1822$-$37 with the 15-40 keV HXD-PIN data. Abscissa shows period difference from a fiducial value of 0.5924337 s. (b) The 15-40 keV pulse profiles, folded at the fiducial period in (a). The background is included, at a level of 15\% of the total counts.}\label{efsearch}
\end{figure*}

We also tried to detect the pulsation at a period $\sim$0.59 s \citep{2001ApJ...553L..43J}, using only the HXD-PIN data because of the low time resolution of the XIS.
After applying barycentric corrections to the individual HXD-PIN events,
we further corrected the event arrive times for the expected orbital delay, $\Delta t$, in 4U 1822$-$37.
This $\Delta t$ is calculated with the pulsar's semi-major axis $a$ and the inclination $i$ as 
\begin{equation}
\Delta t = \frac{a\sin i}{c} \sin\left[2\pi\times\left(\frac{t}{P_{\rm orb}}-\phi_{0}\right)\right],
\end{equation}
where $\phi_{0}$ (0 $\leq \phi_{0} \leq$ 1) is the initial orbital phase.
We fixed $a\sin i$ at 1.006 lt-s, after an accurate measurement by \citet{2001ApJ...553L..43J}, and chose $\phi_{0}=0.445$ so that $\Delta t$ becomes maximum at the observed X-ray dips.
%Jonker et al. (2001) reported $(a\sin i)/c$=1.006$\pm0.005$ and we fixed $(a\sin i)/c$ as 1.006 and corrected events using the equation.
After these corrections, we calculated a 15-40 keV periodogram.
As shown in figure \ref{efsearch}(a), the pulsation was detected with a high ($>$ 99\% confidence) significance at a period of $P_{\rm s}$ = 0.5924337 $\pm$ 0.000001 s.
The pulse profile folded at $P_{\rm s}$ is presented in figure \ref{efsearch}(b).
As already reported \citep{2001ApJ...553L..43J}, the pulse fraction is rather small, $\sim\pm 5$\% in the relative peak amplitude.

\begin{figure}
  \begin{center}
    \includegraphics[width=80mm,height=80mm]{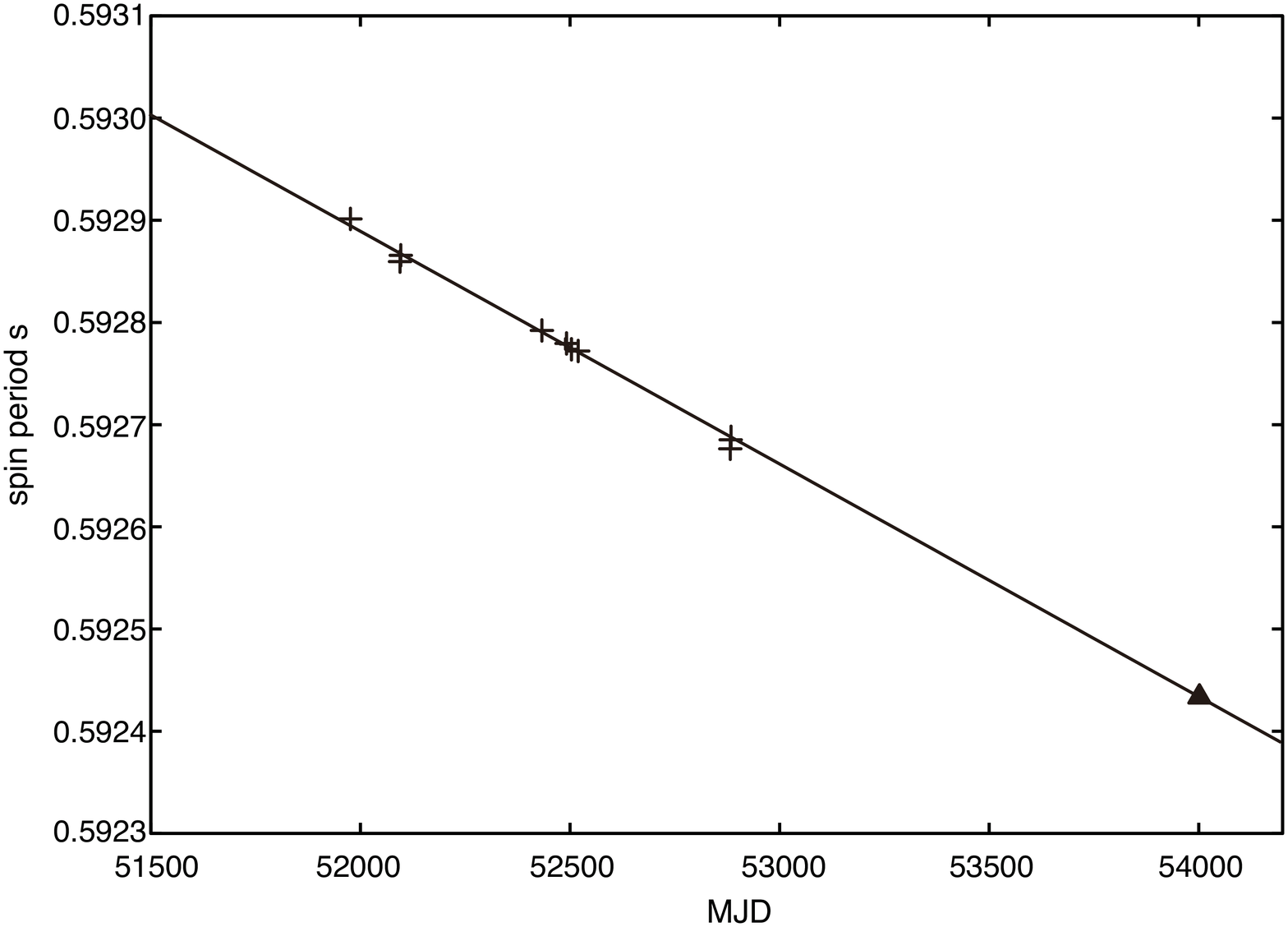}
    %%% \FigureFile(width,height){filename}
  \end{center}
  \caption{The long-term history of $P_{\rm s}$ of 4U 1822$-$37 based on \citet{2010MNRAS.409..755J} and the present work. The triangle indicates the present result.}\label{spinhistory}
\end{figure}

In figure \ref{spinhistory}, the measured value of $P_{\rm s}$ is compared with previous pulse-period measurements.
Over the past $\sim$ 6 years, the object has thus been spinning up monotonically 
with an approximately constant rate of $\dot{P_{\rm s}}$$=-$2.43$\pm$0.05$\times$10$^{-12}$ s s$^{-1}$, or $P_{\rm s}/\dot{P_{\rm s}}$=6.7 kyr.
This value of $\dot{P_{\rm s}}$ reconfirms the previous measurements by \citet{2010MNRAS.409..755J}.
%we corrected barycentric coordinate with ``aebarycen".
%At a folding time 0.592412, we detected pulsation over 99\% and shows pulse profile in the Figure.3.

%\begin{figure}
%\begin{center}
%\epsscale{0.8}
%\plotone{figures/lc_xis_pin_forpaper.eps}
%\caption{Background subtracted light curves of 4U 1822-37 with 128 s bins, obtained with XIS0+XIS3 (top panel; 1-10 keV) and HXD-PIN (middle panel; 15-60 keV).
%The bottom panel shows the HXD-PIN vs XIS hardness ratio.}
%\label{lightcurve}
%\end{center}
%\end{figure}

%\begin{figure}
%\begin{center}
%\epsscale{0.8}
%\plotone{figures/folded_orbitalperiod_xis_pin_ratio.eps}
%\caption{The same light curves of 4U 1822-37 as in figure1, folded at the orbital period of 20054 s. The three panels correspond to those of figure 1.}
%\label{foldedlightcurve}
%\end{center}
%\end{figure}

%\begin{figure}
%\begin{center}
%\includegraphics[angle=270,scale=0.3,angle=90]{figures/efsearch_pulseprofile_forpaper_2.eps}
%\caption{(a) : The periodogram of 4U 1822-37 with the 15-30 keV HXD-PIN data. (b) The 15-30 keV pulse profiles, folded at the pulse period of 0.5924337 s.
%The background is included 15\% level.}
%\label{lightcurve}
%\end{center}
%\end{figure}

%\begin{figure}
%\begin{center}
%\epsscale{0.8}
%\plotone{figures/spin_evol_mod.eps}
%\caption{The long-term history of $P_{\rm s}$ of 4U 1822-37 based on Jain et al. (2010) and the present work.}
%\label{lightcurve}
%\end{center}
%\end{figure}

\subsection{Spectral analysis}
\begin{figure*}
  \begin{center}
    \includegraphics[width=160mm,height=80mm]{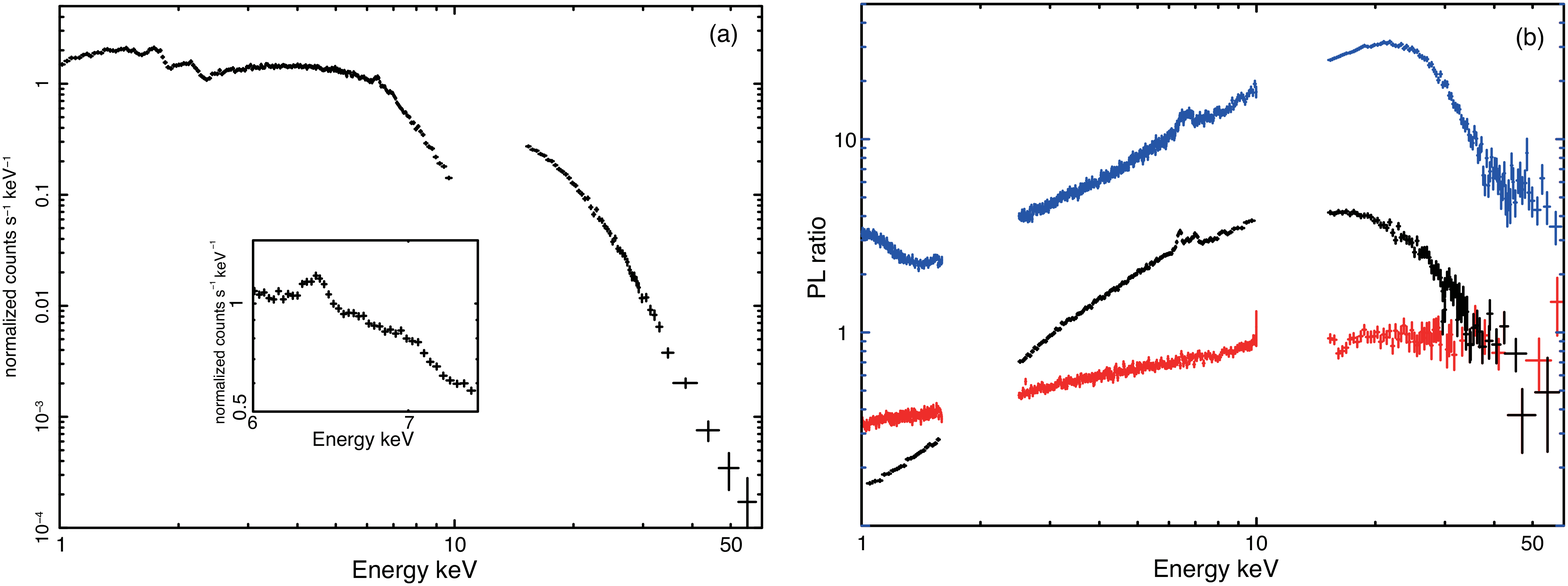}
    %%% \FigureFile(width,height){filename}
  \end{center}
  \caption{(a) Background subtracted spectra of 4U 1822$-$37. The inset shows an expanded view around the iron lines. (b) Suzaku spectra of 4U 1822$-$37 (black; same as panel a), Her X-1 (blue), and EXO0748$-$676 (red), all normalized by a common power-law of photon index 2.0.}\label{compsepctra}
\end{figure*}

Figure \ref{compsepctra}(a) shows 1-50 keV spectra of 4U 1822$-$37 obtained with the XIS and HXD-PIN.
%Thus, an apparently narrow Fe$_{\rm K\rm-\alpha}$ line is observed at $\sim$6.4 keV.
The same spectra, normalized to a power-law model of photon index 2.0, are compared in figure \ref{compsepctra}(b) with 
those of another dipping source EXO 0748$-$676 and the X-ray pulsar Her X-1.
%We detected Fe$_{\rm K\rm-\alpha}$ at $\sim$ 6.4 clearly.
%We compared the spectra among another dipping source EXO 0748-676 and strong magnetic field pulsar Her X-1 in Figure 5 (b).
%Between the spectra of 4U 1822-37 and those of EXO 0748-676, there are some difference.
The spectral shape of 4U 1822$-$37 is similar to that of Her X$-$1 while different from that of EXO 0748$-$676, in the following two points.
First, the spectrum of 4U 1822$-$37 shows a very hard slope (with a power-law photon index of $\Gamma\sim$1) below 10 keV and a steep cutoff at $\sim$ 15 keV, 
in resemblance to the Her X-1 spectrum, while that of EXO 0748$-$676 exhibits a softer ($\Gamma\sim$2) slope without clear high-energy cutoff.
% like Her X-1.%EXO 0748-676 showed softer ($\Gamma\sim$2) and no cutoff spectra.
Second, complex Fe ${\rm K}$-${\rm\alpha}$ lines are seen at $\sim$ 6.4 keV in the spectra of 4U 1822$-$37 and Her X-1,
while they are absent in the case of EXO 0748$-$676.
These suggest that the NS in 4U 1822$-$37 has strong magnetic fields, like that in Her X$-$1 \citep{2008PASJ...60S..57E}.

%As Parmar et al. 2000 have already reported, 4U 1822-37 showed much harder spectra than 
%other dipping source spectra or LMXB like EXO 0748-676.
%There are is iron line and a cutoff  on the EXO 0748-676 spectra.
%However, 4U 1822-37 and Her X-1 showed the Fe$_{\rm K\rm-\alpha}$ and cutoff $\sim$ 10 keV.
%From the comparison among them, 4U 1822-37 has similar feature to Her X-1 and we can consider 4U 1822-37 has strong magnetic fields.

\begin{figure}
  \begin{center}
    \includegraphics[width=80mm,height=80mm]{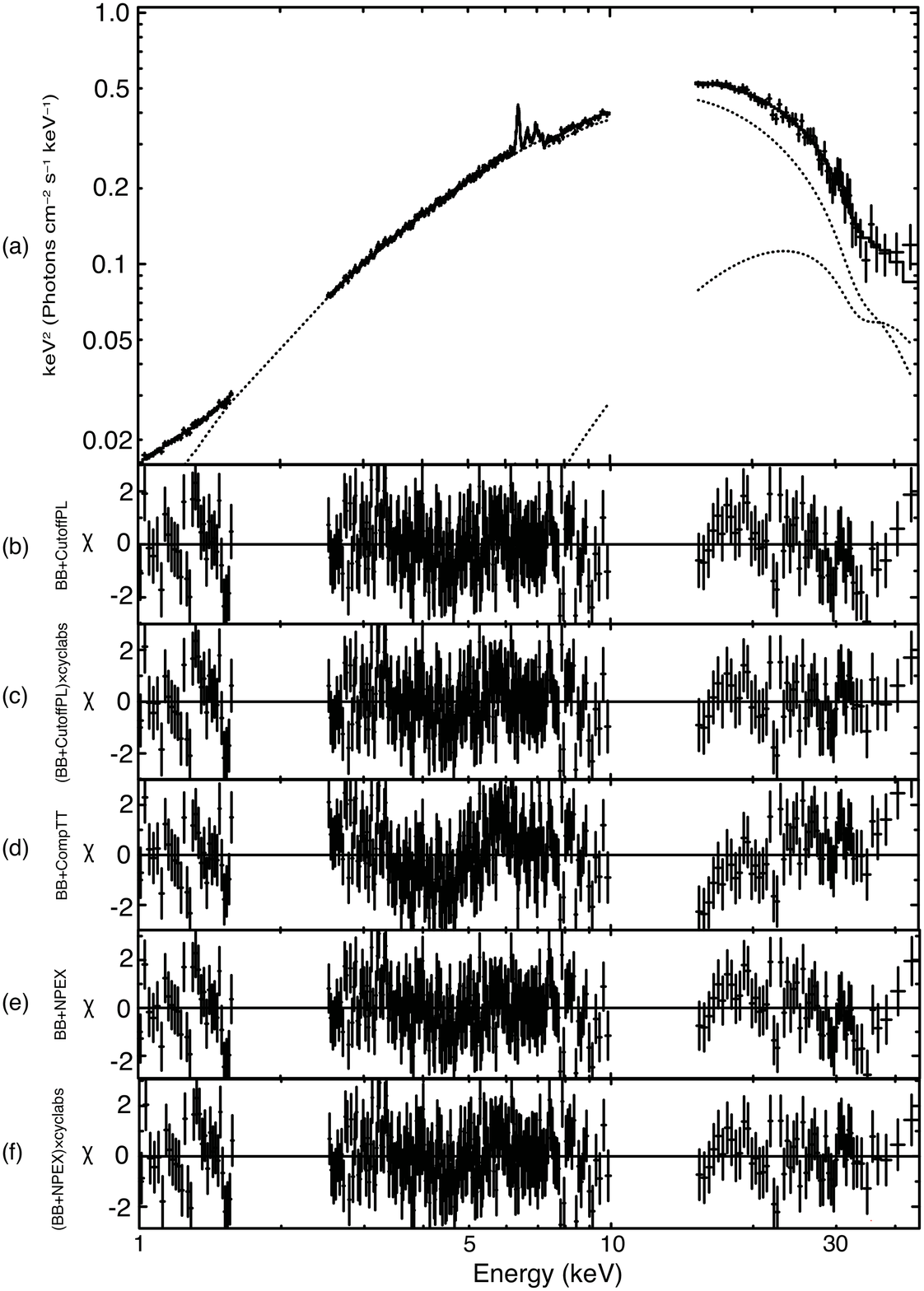}
    %%% \FigureFile(width,height){filename}
  \end{center}
  \caption{(a) $\nu F\nu$ spectra of 4U 1822$-$37, fitted with wabs$\times$(BB+NPEX+3$\times$Gaussian)$\times$cyclabs. The softer and harder continuum components represent the negative and positive power-laws of the NPEX model, with a common exponential cutoff factor. (b) Fit residuals with the wabs$\times$(BB+CutoffPL+3$\times$Gaussian) model. (c) Those with wabs$\times$(BB+CutoffPL+3$\times$Gaussian)$\times$cyclabs. (d) Those with wabs$\times$(BB+CompTT+3$\times$Gaussian). (e) Those with wabs$\times$(BB+NPEX+3$\times$Gaussian). (f) Those with wabs$\times$(BB+NPEX+3$\times$Gaussian)$\times$cyclabs.}\label{spectrafit}
\end{figure}

To quantify the XIS and HXD-PIN spectra of 4U 1822$-$37, 
we fitted them jointly with an absorbed cutoff-power-law (CutoffPL) model, including
three narrow Gaussians to represent Fe-K$\alpha$, Fe XXV K$\alpha$, and Fe XXVI K$\alpha$ lines, at $\sim$6.4, $\sim$6.7 and $\sim$6.9 keV, respectively.
%(with the $\sigma_{K\alpha}$=0.04 keV and $\sigma_{\rm Fe-XXV,Fe-XXVI}$=0.05 keV)
The photon index $\Gamma$, the cutoff energy $E_{\rm cut}$, and normalization of CutoffPL were left free, as well as the column density of absorption.
Similarly, the three Gaussians were allowed to have free center energies, free normalizations, and free widths.
% and the center energy and normalizations of the Gaussians.
%The lowest-energy Gaussian, representing neutral Fe-K$\alpha$ line, was constrained to have a width of $\sigma=0.04$ keV, 
%while the other two (Fe XXV K$\alpha$ and Fe XXVI K$\alpha$) $\sigma=0.05$ keV. 
The HXD vs. XIS cross normalization was fixed at 1.16 \citep{2007PASJ...59S..53K}.
%we fitted them with five models, namely,
%BlackBody (BB)+Thermal Comptonization (comptt) (Titarchuck et al. 1994),
%BB+Cutoff Power-law (cutoffpl),
%BB+cutoffpl+CyclotronAbsroption (cyclabs),
%BB+Negative-Positive with Exponential Cutoff (NPEX),
%and BB+NPEX+Cyclabs, and obtained the results shown in table 1.
%As we showed in figure 6., we detected three iron lines and edge at $\sim$ 7.2 keV.
%Three gaussians and edge were included in the all models for presenting lines and edge of iron.
%We fitted them with very simple model wabs$\times$(cutoffpl + 3gaussians).
However, a large data excess below 3 keV made the fit unacceptable, with a reduced chi-square of $\chi^{2}_{\nu}>$3.0 for $\nu$=287 degrees of freedom.
To account for this ``soft excess'' feature, we chose a blackbody ({\tt BB}), and fitted the data with {\tt wabs$\times$(BB+CutoffPL+3gaussians)}.
%There was a large excess below 3 keV and a reduced chi squared $\chi^{2}$ is over 3 with a degree of freedom 287.
%Replacing cutoffpl to other models, the results were nearly the same and not acceptable.
%For the soft excess, we chose BB and fitted with {\tt wabs$\times$(BB+CutoffPL+3gaussians)}.
A {\tt BB} with a temperature of $\sim$ 0.1 keV improved the fit to $\chi^{2}_{\nu}$=1.21 ($\nu$=284).
%including the black body (temperature $\sim$ 0.1 keV), the reduced chi square of this model was much better than that of the previous model.
%From this results, we thought that the black body is necessary at the low energy.
However, as shown in figure \ref{spectrafit}(b), there still remained negative residuals at $\sim$ 35 keV.
Since they are suggestive of a CRSF, we employed a model of the form of {\tt wabs$\times$(BB+CutoffPL+3gaussians)$\times$cyclabs}, where {\tt cyclabs} represents 
the cyclotron resonance absorption factor given, e.g., in \citet{1999ApJ...525..978M}.
%error between data and model at $\sim$ 35 keV in the middle panel of Figure 6.
%To fit the residual error, we used a model wabs$\times$(BB+cutoffpl+3gaussians)$\times$cyclabs and fit was improved.
By allowing the resonance energy, width, and the depth of {\tt cyclabs} to vary freely,
the fit has been improved to $\chi^{2}_{\nu}=1.14$ for $\nu$=280.
%the fit has been improved by $\Delta\chi^{2}=-25.9$  for $\Delta\nu=-3$, and become acceptable ($\chi^2_{\nu}=1.14$ for $\nu$=280) with a null hypothesis probability of 5.4\%.
As summarized in table 1, the center energy of CRSF was obtained as $E_{\rm CRSF}$=33$\pm$2 keV.
%the depth of absorption is $D$=0.4$^{+0.6}_{-0.2}$, and the resonance width is constrained as $W<$ 7 keV.

To make the model more physical, we once excluded the cyclabs factor, and tried a thermal Comptonizetion ({\tt CompTT}; e.g., \cite{1995ApJ...450..876T}) and a 
``negative-positive power-law with exponential cutoff" ({\tt NPEX}; e.g., \cite{1999ApJ...525..978M}) continua instead of {\tt CutoffPL}, while retaining the {\tt BB} component and the three Gaussians.
%Using the wabs$\times$(BB+CompTT+3gaussians), the residual error was much larger than the case of cutoffpl and it is not acceptable.
%The NPEX is more suitable model for the fitting because $\chi^{2}_{\nu}$ of the model is smaller than that of cutoffpl.
As summarized in table 1, the {\tt CompTT} continuum was less successful than that with {\tt CutoffPL}, while the {\tt NPEX} continuum was better.
As shown in figure \ref{spectrafit}(e), however, the {\tt BB+NPEX} continuum again left the negative residuals at $\sim$35 keV like the case of {\tt CutoffPL}.
We hence revived the {\tt cyclabs} factor, and fitted the data with {\tt wabs$\times$(BB+NPEX+3gaussian)$\times$cyclabs}.
The fit has become fully acceptable, $\chi^{2}_{\nu}$=1.13 with $\nu$=279.

The obtained {\tt cyclabs} parameters, shown in table 1, are not much different from those with the {\tt CutoffPL} continuum case,
except that the width has now been constrained as 5.0$^{+5.0}_{-3.0}$.
%As the depth of {\tt cyclabs} is fixed zero, the fit has been changed by $\Delta\chi^{2}=+28.84$ and $\Delta\nu=1$, and showed the depth is positive over 99\%.
While the error ranges in table 1 refer to 90\% confidence limit,
the {\tt cyclabs} depth still remains positive, $D\geq$0.24, if we employ more conservatively 99\% limit (i.e., $\Delta\chi^{2}$=6.63 for a single parameter).
Since the inclusion of the {\tt cyclabs} factor improved the fit by $\Delta\chi^2$=$-$13.7 for $\Delta\nu$=$-$3, or $\Delta\chi^2$/$\Delta\nu$ = 4.56,
an F-test indicates that the fit improvement is significant at a confidence level of 99\%.
Using the {\tt gabs} model for the CRSF instead of the {\tt cyclabs} model gave nearly the same $\chi^{2}_{\nu}$ and a consistent value of $E_{\rm CRSF}$ 
%With the F-test, we checked significant level of the CRSF and got significancy over 99\%.
%Since the depth of {\tt cyclabs} is significantly positive a the 99\% confidence level, 
From these results, we can claim the detection of a CRSF.
%The depth of {\tt cyclabs} is significant within 90\% error and we detected the CRSF over 90\%.
%An F-test between wabs$\times$(BB+NPEX+3gaussian) and wabs$\times$(BB+NPEX+3gaussian)$\times$cyclabs
%indicates that the CRSF factor is significant at a confidence level of $>$ 1$\sigma$.
%we tried F-test to check a significance of CRSF.
%By the F-test, the significance is over 1$\sigma$.

The 1-50 keV flux of this source becomes 1.5$\times10^{-9}$ erg s$^{-1}$ cm$^{-2}$, and 
the luminosity is $L_{\rm X}$ =1.0$\times 10^{36}$ erg s$^{-1}$ at the distance of 2.5 kpc \citep{1982MNRAS.200..793M}.
These are consistent with previous reports \citep{1982MNRAS.200..793M,2001ApJ...557...24I}.

\begin{figure}
  \begin{center}
    \includegraphics[width=80mm,height=80mm]{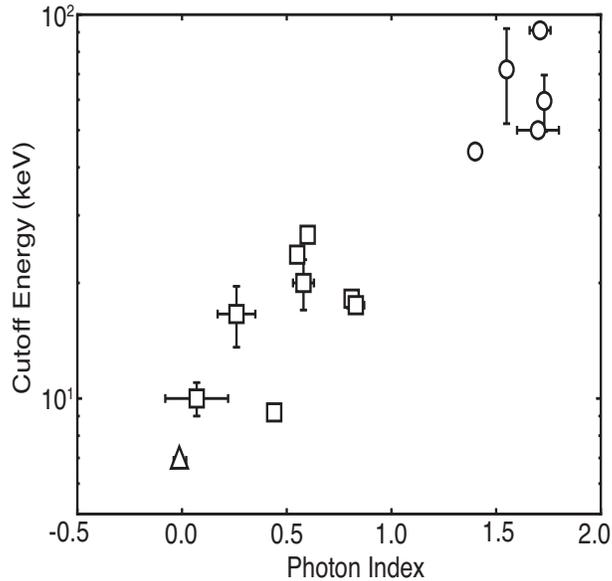}
    %%% \FigureFile(width,height){filename}
  \end{center}
  \caption{Photon indices and cutoff energies of the Suzaku spectra of representative neutron-star binaries. Circles represent hard state LMXBs, squares accreting pulsars, and the triangle 4U 1822$-$37.}\label{gammacutoff}
\end{figure}

\section{Discussion}

%With the results of the timing and spectral analysis, we discuss the magnetic field strength of 4U 1822-37.

\subsection{Spectral shape}
In figure \ref{compsepctra}(b), the spectra of 4U 1822$-$37 were visually compared with those of Her X-1 and EXO 0748$-$676.
For doing this more quantitatively,
%For comparing among the spectral shape more quantitatively, 
%we adopted the $\Gamma$ and $E_{\rm cut}$ as parameters representing spectral shape and analyzed
we fitted the {\tt CutoffPL} model to Suzaku spectra of other hard-state LMXBs (Aql X$-$1, EXO 0748$-$676, XB 1323$-$619, 4U 1636+536 and 4U 1608$-$52), 
and accreting pulsars which have strong magnetic fields (4U 1626$-$67, Cen X-3, SMC X-1, Her X-1, GRO J1008$-$57, 1A 1118$-$61, 1A 0535+26 and LMC X-4).
Figure \ref{gammacutoff} summarized those fit results, on a plane of $\Gamma$ vs $E_{\rm cut}$.
The hard state LMXBs typically have softer ($\Gamma\sim$1.7) slopes and higher cutoff energies ($E_{\rm cut} >$ 50 keV),
while the accreting pulsars show harder ($\Gamma\sim$0.6) slopes together with lower cutoff energies ($E_{\rm}\sim$10 keV).
Clearly, 4U 1822$-$37 is more similar to the accreting pulsars than to the hard-state LMXBs.
In terms of the continuum slope, 4U 1822$-$37 is thus inferred to have common characteristics of accreting NSs with $B\sim10^{12}$ G.

\subsection{The Cyclotron Resonance Scattering Feature}
As argued in \citet{1999ApJ...525..978M}, the high-energy cutoff of the spectra of accreting pulsars, 
which are often steeper than a thermal rollover as pointed out by \citet{1986LNP...255..198T}, 
is presumably caused by the presence of a CRSF, which is empirically thought to appear at energies of $E_{\rm CRSF}\sim1.5\times E^{1.5}_{\rm cut}$.
Then the NPEX fit results, $E_{\rm cut}$=4.8-6.2 keV, predict a CRSF to appear at $E_{\rm CRSF}$=16-23 keV.
This argument was already employed by \citet{2001ApJ...553L..43J} to suggest $B\sim(1-5)\times10^{12}$ G.

Indeed, as expected, we have detected a CRSF at $E_{\rm CRSF}\simeq33$ keV with a high significance. 
Using the basic relation of $B$=$(1+z)\times(E_{\rm CRSF}/11.6)\times10^{12}$ G (e.g., \cite{1999ApJ...525..978M}),
where $z$ is the gravitational redshift of the NS,
we obtain $B$=(2.8$\pm$0.2)$\times10^{12}$ G assuming $z$=0 for simplicity.
This gives the most convincing evidence that the NS in 4U 1822$-$37 is strongly magnetized.

Just to make the CRSF detection more convincing, let us examine the derived CRSF parameters. 
As already confirmed, the resonance energy $E_{\rm CRSF}$ is consistent with the continuum shape. 
The depth, $D\sim0.4$, is reasonable, in comparison with typical values of $D$=0.1-1.7 found in other objects. 
In addition, $W$=2-10 keV=(0.06-0.30)$E_{\rm CRSF}$ agrees, within errors, with the general scaling of $W$=(0.27-0.50)$E_{\rm CRSF}$ found by \citet{1999ApJ...525..978M}.
Thus, the present CRSF interpretation of the $\sim$33 keV spectral feature is considered reasonable.
%We detected clear evidence of the strong magnetic fields, CRSF at 33 keV over 99\% confidence.
%Using the cutoff energy, the energy of CRSF is also estimated with $E_{\rm CRSF}$=1.48$\cdot E^{1.5}_{\rm cut}$ (Makishima et al. 1999).
%As we show in table 1, $E_{\rm cut}$=4.8-7.1 keV and we derives $E_{\rm CRSF}$=15-28 keV.
%The energy of CRSF which we estimated from $E_{\rm cut}$ reproduces the result Jonker et al. 2001 reported and is consistent with the energy we detected within 20\%.
%We consider the relation between $E_{\rm CRSF}$ and the width of CRSF,  Width$\sim$(0.27-0.5)$E_{\rm CRSF}$  (Makishima et al. 1999).
%Using the relation, the width is estimated as 8.9-16.5 and the estimation result is consistent with the fitted result within errors.
%The features of the CRSF we detected are consistent with those of other objects CRSF.
%The magnetic field strength $B_{12}$ united by 10$^{12}$ G is represented by the energy of the CRSF
%\begin{equation}
%(B_{12}) = (1+z)\times (E_{\rm CRSF}/11.6)
%\end{equation}
%z means the red shift parameter by gravity of the NS and we assume z=0 in this time.
%From the $E_{\rm CRSF}=33$ keV, we derived $B_{12}$=2.8 and the magnetic field strength is $2.8\times10^{12}$ G.
%We conclude that 4U 1822-37 has strong magnetic field, 2.8$\times10^{12}$ G.

\subsection{Spin period and Spin-up rate}
The present HXD data yielded $P_{\rm s}$ = 0.5924337 s and $\dot{P_{\rm s}}$ = -2.43$\times10^{-12}$ s s$^{-1}$.
%From our analysis, we measured the spin p$P_{\rm s}$ = 0.592398 s and spin up rate $\dot{P_{\rm s}}$ = -2.85$\times10^{-12}$ ss$^{-1}$.
When these values, together with $B$=2.8$\times10^{12}$ G, are substituted into the accretion torque formula by \citet{1978ApJ...223L..83G}, namely 
\begin{equation}
%B = 0.12\times10^{12}\times(-\dot{P_{\rm s}}\cdot10^{12})^{7/2}\times P_{\rm s}^{7}\cdot(L/10^{37})^{3}
-\dot{P} = 1.84\times10^{-12}\times B_{12}^{2/7}\times(PL_{37}^{3/7})^{2} {\rm s\cdot s^{-1}}
\end{equation}
where $L_{37}$ is the luminosity in units of 10$^{37}$ erg s$^{-1}$ and $B_{12}$ is the magnetic field in $10^{12}$ G, we obtain $L\sim3\times10^{37}$ erg s$^{-1}$.

%Using the values obtained by this analysis, $P_{\rm s}$, $\dot{P_{\rm s}}$ and $B_{12}$, we could estimated the luminosity as 4$\times10^{37}$ erg s$^{-1}$.
This luminosity is much higher than the value of $\sim$1.0$\times10^{36}$ erg s$^{-1}$ derived from the observed flux and an assumed distance of 2.5 kpc \citep{1982MNRAS.200..793M}.
One possible cause of this discrepancy is that the object is in really located at $\sim10$ kpc distance, instead of the 2.5 kpc which is based on some assumptions.
An alternative possibility, already pointed out previously (\cite{1982ApJ...257..318W,2000A&A...356..175P}), 
is that the object appears unusually X-ray faint due, e.g., 
%to X-ray scattering by some coronae above the accretion disk.
to X-ray obscuration by some ionized materials on the accretion disk.
Indeed, its X-ray to optical luminosity ratio of $\sim$20 for $L_{\rm X}\sim$1$\times$10$^{36}$ erg s$^{-1}$ (\cite{1982MNRAS.200..793M}) 
is much lower than a typical value of $\sim$500 for LMXBs (\cite{1995xrbi.nasa...58V}).
The ratio of 4U 1822$-$37 will increase to $\sim$600 if we employ $L_{\rm X}\sim$3$\times$10$^{37}$ erg s$^{-1}$.
%This discrepancy has already been discussed scattering by an accretion disk corona \citep{1982ApJ...257..318W,2000A&A...356..175P}.
%As another possibility, some assumptions were used to derive the distance and the luminosity has uncertainty.
%Considering a relation between the X-ray luminosity and the optical luminosity $L_{\rm opt}$, a typical ratio between $L$ and $L_{\rm opt}$ of LMXB is $\sim$500 \citep{1995xrbi.nasa...58V}.
%In the case of this source, the ratio is $\sim$ 20 with $L_{\rm X}$=1$\times10^{36}$ erg s$^{-1}$ and $\sim$600 adopting $L_{\rm X}$=$3\times10^{37}$ erg s$^{-1}$.
%The luminosity derived from this work, 3$\times10^{37}$ erg s$^{-1}$, is more adequate value to that of 4U 1822-37.
%From the luminosity of Her X-1 $\sim10^{37-38}$ (Enoto et al. 2008, Ji et al. 2009), 
%we prefer the higher result, $L=4\times10^{37}$ erg s$^{-1}$.

\subsection{Iron lines}
Another interesting discussion may be performed on the iron lines in the spectra.
%Another empirical comparison may be conducted on iron lines in the spectra. 
Generally, LMXBs have weak, sometimes broad \citep{2008ApJ...674..415C} iron lines with small equivalent width (EWs); e.g., $20\pm15$ eV \citep{2012PASJ...64...72S}.
Strongly magnetized NSs, including Her X$-$1 and GX 1+4, in contrast show narrow iron lines with significantly larger EWs ($>$ 50 eV; \cite{2010ApJ...715..947T}).
%These differences can be employed as an empirical way of estimating $B$ of an NS, although their theoretical account is beyond the scope of the present paper.
These differences can be employed as an additional empirical argument to strengthen the high $B$ scenario of 4U 1822$-$37,
although theoretical account of this issue is beyond the scope of the present paper.
%As a possibility, Accretion geometris were changed by the magnetic field strength and the $EW$ of iron lines becomes larger.
%In this respect, 4U 1822-37 has a narrow ($\sigma$=0.04 keV) and strong iron line (EW$\sim$ 50 eV), again similar to typical X-ray pulsar.
In this respect, The narrow ($\sigma$=0.04 keV) and strong (EW$\sim$ 50 eV) Fe K$\alpha$ line, detected at 6.4 keV with the XIS, clearly classifies this NS into the high-field category.
% again similar to typical X-ray pulsar.

To be somewhat more quantitative, the width of the Fe K$\alpha$ line, 0.04$\pm$0.02 keV, implies that the velocity $v_{\rm gas}$ of the gases which emit the iron line is $\sim$ 0.6\% of the light speed.
Assuming that the gas obeys Keplar rotation, its distance becomes $r_{\rm gas} = GM/v^2_{\rm gas}\sim 5.7\times10^9$ cm ($G$ being the gravitational constant and $M\sim1.4M_{\odot}$ the NS mass),
which is comparable to the Alfven radius ($\sim 10^{8-9}$ cm) for $B\sim10^{12}$ G while much larger than the NS radius ($\sim10^{6}$ cm).
Therefore, the matter creating the fluorescent Fe-K line is likely to be stored on the Alfven surface.

In the present data, we detected not only the nearly neutral Fe K$\alpha$ line, but also the other two lines 
which are identified as Helium like (Fe XXV) line and Hydrogen like iron (Fe XXVI) line as judged from their line energies.
Empirically, such ionized iron lines are observed from X-ray pulsars mainly when the source luminosity is high; e.g., $L_{\rm X}\geq$1$\times10^{37}$ erg s$^{-1}$. 
Such examples include Cen X-3 \citep{2011ApJ...737...79N}, Her X-1 \citep{2009ApJ...700..977J}, LMC X-4 \citep{2009ApJ...696..182N}, and Be pulsars in luminous outbursts \citep{2013ApJ...764..158N}.
Quantitatively this is reasonable, because ionization of circum-source materials is determined by so-called ionization parameter $\xi$=$L_{\rm X}/(n_{\rm e}\cdot r^{2})$,
where $n_{\rm e}$ is the electron density of the line emitter while $r$ is its distance from the NS (ionizing photon source).
Assuming that the two ionized lines in 4U 1822$-$37 are emitted by the same material that is different from those emitting the 6.4 keV line,
we estimate as $\xi\sim2000$, from \citet{1982ApJS...50..263K} and the comparable EWs of the two ionized lines (table1).
%its ionization parameter $\xi$ can be calculated from the ratio between the EW of Fe XXV$ and that of Fe XXVI.
%As the ratio is $\sim 1$, $\xi$ should be $\sim 2000$ by according to Kallman and McCary (1987).
Employing the same argument as for $r_{\rm gas}$, we may obtain $r\sim3\times10^{9}$ cm from the line width of $\sim0.05$ keV. 
Adopting $n_{\rm e}$=$5\times10^{15}$ cm$^{-3}$ from \citet{2013A&A...549A..33I},
the luminosity is 1$\times10^{37}$ erg s$^{-1}$, in agreement with estimate using equation (2).
Thus, the source is considered to be relatively luminous in X-rays.

During the net exposure of 37 ks ($\sim10 $h), no X-ray bursts were detected. 
Similarly, there have been not previous reports of burst detection from this objects. 
While this could be due to rather infrequent burst occurrence at $L_{\rm X}>2\times10^{37}$ erg s$^{-1}$ (\cite{2003A&A...405.1033C}), 
a more appropriate explanation would be to regard 4U 1822$-$37 as an X-ray pulsar, which do not produce X-ray bursts.

We hence conclude that 4U 1822$-$37 is another example of LMXB that contains a strongly magnetized ($B\sim$3$\times$10$^{12}$ G) NS. 
Furthermore, its luminosity is likely to be $\sim$3$\times$10$^{37}$ erg s$^{-1}$, instead of the previously reported value of $\sim$1$\times$10$^{36}$ erg s$^{-1}$,
although it is at present unclear whether this discrepancy is due to inaccurate distance estimate, on due to the edge-on source geometry which could reduce the X-ray flux reaching us.

We thank all members of the Suzaku hardware and software teams and the Science Working Group. 
M.S., K.M. and T.E. are supported by the Japan Society for the Promotion of Science (JSPS) Research Fellowship for Young Scientists, the Grant-in-Aid for Scientific Research (A) (23244024) from JSPS, and Grant-in-Aid for JSPS Fellows, 24-3320, respectively.

\begin{center}
%\begin{longtable}{ccccccc}
\begin{longtable}{ccccccc}
  \caption{Results of the model fit to the 1-50 keV {\it Suzaku} spectra of 4U 1822$-$37.}\label{tab:LTsample}
  \hline              
  Components & Parametars & BB+cutoffPL & BB+cutoffPL & BB+CompTT & BB+NPEX & BB+NPEX \\ 
   		       &                     &                      &     +Cyclabs         &                       &                  &		+Cyclabs\\
\endfirsthead
  \hline  
%  name & value & value2 & test \\
%  Components & Parametars & BB+cutoffPL & BB+cutoffPL+Cyclabs & BB+CompTT & BB+NPEX & BB+NPEX+Cyclabs \\ 
\endhead
  \hline
\endfoot
  \hline
\endlastfoot
  \hline
  wabs&$N_{\rm H}$ (10$^{22}$ cm$^{-2}$)& 0.27$\pm$0.06 & 0.33$\pm$0.07 & 0.26$\pm$0.07 & 0.28$\pm$0.07 &0.29$^{+0.07}_{-0.08}$ \\
\hline
BB&$kT$ (keV) & 0.15$\pm$0.01 & 0.15$\pm$0.01 & 0.15$\pm$0.01 & 0.16$\pm$0.01 & 0.15$\pm$0.01 \\
\hline
CompTT&$kT_{\rm in}$ (keV) & --- &--- & 1.00$\pm$0.02 & ---  & --- \\
&$kT_{\rm e}$ (keV) & --- & --- & 4.39$\pm$0.07 & --- & --- \\
&$\tau_{\rm comp}$ & --- & --- & 7.6$\pm$0.2 & --- & --- \\
\hline
NPEX&$E_{\rm cut}$ (keV) & 6.9$\pm$0.1& 7.1$\pm$0.2 & --- & 4.8$\pm$0.2 & 6.2$\pm$1.0 \\
&$\Gamma$ & -0.01$\pm$0.03 & 0.03 $\pm$ 0.03 & --- & -0.12$\pm$0.03 & -0.04$^{+0.08}_{-0.06}$ \\
\hline
edge&$E_{\rm edge}$ (keV) & 7.23$_{-0.05}^{+0.04}$ & 7.23$_{-0.05}^{+0.04}$ & 7.2$\pm$0.1 & 7.2$\pm$0.1 & 7.23$^{+0.05}_{-0.06}$\\
&$\tau$ & 0.10$\pm$0.01& 0.09$\pm$0.01 & 0.06$\pm$0.02 & 0.09$\pm$0.01 & 0.09$\pm$0.02 \\
\hline
Gaussian&$E_{\rm Fe K\alpha}$ (keV)& 6.39$\pm$0.01 & 6.39$\pm$0.01 & 6.38$\pm$0.01& 6.39$\pm$0.01 & 6.39$\pm$0.01\\
&$\sigma_{\rm Fe K\alpha}$ (keV)& 0.04$\pm$ 0.02 & 0.04$\pm$ 0.02 & 0.04$\pm$ 0.02 & 0.04$\pm$ 0.02 & 0.04$\pm$ 0.02\\
&EW$_{\rm Fe K\alpha}$ (eV) & 43$_{-2}^{+3}$ & 48$_{-4}^{+5}$ & 50$\pm$2 & 49$_{-3}^{+7}$  & 49$_{-4}^{+3}$\\
\hline
Gaussian&$E_{\rm Fe XXV}$  (keV)& 6.66$\pm$0.03 & 6.68 $\pm$0.04 & 6.69$\pm$0.03  & 6.68 $\pm$0.04  & 6.68 $\pm$0.04\\
&$\sigma_{\rm Fe XXV}$ (keV) & 0.05$\pm$0.03 & 0.05$\pm$0.03 & 0.05$\pm$0.03 & 0.05$\pm$0.03 & 0.05$\pm$0.03 \\
&EW$_{\rm Fe XXV}$ (eV)& 11$\pm$3 & 16$_{-5}^{+3}$ & 20$\pm$3  & 14$\pm$2 & 13$_{-1}^{+5}$\\
\hline
Gaussian&$E_{\rm Fe XXVI}$ (keV) & 6.96$\pm$0.04 & 6.97$\pm$0.03 & 6.98$_{-0.03}^{+0.05}$ & 6.98$\pm$0.03 & 6.97$\pm$0.03\\
&$\sigma_{\rm Fe XXVI}$ (keV) & 0.05$\pm$0.03 & 0.05$\pm$0.03 & 0.05$\pm$0.03 & 0.05$\pm$0.03 & 0.05$\pm$0.03 \\
&EW$_{\rm Fe XXVI}$ (eV) & 13$\pm$3 & 15$_{-2}^{+3}$ & 18$\pm$2 & 13$_{-5}^{+1}$ & 16$^{+3}_{-4}$\\
\hline
CRSF&$E_{\rm CRSF}$ (keV) & --- & 33$\pm$2 & --- & --- & 33$\pm$2\\
&$D$  & --- & 0.4$^{+0.6}_{-0.2}$& --- & --- & 0.4$_{-0.1}^{+0.2}$\\
&$W$ (keV) & --- & $<$ 7.0& --- & --- & 5.0$_{-3.0}^{+5.0}$\\
\hline
fit goodness&$\chi^{2}$ (ndf) & 345.0 (284) & 319.1 (280) & 451.9 (284) & 330.5 (282) & 316.8 (279)\\
\hline
\end{longtable}
\end{center}

%\begin{longtable}{ccccccc}
%  \caption{Sample of ``longtable"}\label{tab:LTsample}
%  \hline              
%  name & value1 & value2 & test \\ 
%\endfirsthead
%  \hline
%  name & value & value2 & test \\
%\endhead
%  \hline
%\endfoot
%  \hline
%\endlastfoot
%  \hline
%  aaaaa & bbbbb & ccccc &\\
%  ...... & ..... & ..... &\\
%  ...... & ..... & ..... &\\
%  ...... & ..... & ..... &\\
%  xxxxx & yyyyy & zzzzz & \\
%\end{longtable}

%\bigskip

%Acknowledgement should be placed at end of main text.
%(NOT after the Appendix.)

%\appendix
%\section{Method of .....}

%\section{Approximation of ...}

%\section*{Complete data}

%%%
% See the manual for the detail.
%%%

\end{document}